\documentclass{mn2e} 
\usepackage{color}
\usepackage{dcolumn}
\usepackage{bm}
\usepackage{epsfig}

\begin{document}

\twocolumn

\title[Cosmic variance] {Quantifying cosmic variance}

\author[Driver \& Robotham]  {Simon P.~Driver$^{1,2}$ and Aaron S.G.~Robotham$^{1}$\\
$^{1}$ Scottish Universities Physics Alliance (SUPA)\\ 
 School of Physics \& Astronomy, University
  of St Andrews, North Haugh, St Andrews, Fife, KY16 9SS, UK\\
$^{2}$ Visiting Professor, International Centre for Radio Astronomy Research (ICRAR), University of Western Australia, Crawley, WA 6009, Australia\\
}

\date{Received XXXX; Accepted XXXX}
\pubyear{2006} \volume{000}
\pagerange{\pageref{firstpage}--\pageref{lastpage}}

\maketitle
\label{firstpage}

\begin{abstract}
We determine an expression for the cosmic variance of any ``normal''
galaxy survey based on examination of $M^* \pm 1$ mag galaxies in the
SDSS DR7 data cube. We find that cosmic variance will depend on a
number of factors principally: total survey volume, survey aspect
ratio, and whether the area surveyed is contiguous or comprised of
independent sight-lines. As a rule of thumb cosmic variance falls
below 10\% once a volume of $10^7h_{0.7}^{-3}$Mpc$^3$ is surveyed for
a single contiguous region with a 1:1 aspect ratio. Cosmic variance
will be lower for higher aspect ratios and/or non-contiguous
surveys. Extrapolating outside our test region we infer that cosmic
variance in the entire SDSS DR7 main survey region is $\sim 7$\% to $z
< 0.1$

The equation obtained from the SDSS DR7 region can be generalised to
estimate the cosmic variance for any density measurement determined
from normal galaxies (e.g., luminosity densities, stellar mass densities and
cosmic star-formation rates) within the volume range $10^3$ to $10^7
h^{-3}_{0.7}$Mpc$^3$.

We apply our equation to show that 2 sightlines are required to ensure
cosmic variance is $<10$\% in any ASKAP galaxy survey (divided into
$\Delta z \sim 0.1$ intervals, i.e., $\sim 1$ Gyr intervals for $z
<0.5$). Likewise 10 MeerKAT sightlines will be required to meet the
same conditions. GAMA, VVDS, and zCOSMOS all suffer less than 10\%
cosmic variance ($\sim$ 3\%-8\%) in $\Delta z$ intervals of 0.1, 0.25,
and 0.5 respectively. Finally we show that cosmic variance is
potentially at the 50-70\% level, or greater, in the HST Ultra Deep
Field depending on assumptions as to the evolution of clustering. 100
or 10 independent sightlines will be required to reduce cosmic
variance to a manageable level ($<10$\%) for HST ACS or HST WFC3
surveys respectively (in $\Delta z \sim 1$ intervals). Cosmic variance
is therefore a significant factor in the $z>6$ HST studies currently
underway.
\end{abstract}


\section{Introduction}
The Universe is not homogeneous except on the largest scales ($>1$Gpc,
Davis et al.~1985). As a consequence number and density measurements
derived from within modest volumes will show greater than Poisson
variation (see Szapudi \& Colombi 1996 for example). This cosmic
variance\footnote{Technically the term sample variance is more correct
  but here we adhere to the current convention of using the term
  cosmic variance to describe perturbations in measurements within our
  Universe due to sampling size.}, or small-scale scale-dependent
inhomogeneity, is often the dominant source of error in many
contemporary extragalactic measurements. Examples include the galaxy
luminosity function/densities (Norberg et al.~2002a; Hill et
al.~2010), the HI mass function (Zwaan et al.~2005), the cosmic
star-formation history (Hopkins \& Beacom~2006), and the stellar mass
density (Wilkins, Trentham \& Hopkins~2008). Generally any number or
density measurement derived from the galaxy population as a whole is
susceptible. Typically, although often neglected in many studies, the
cosmic variance can be estimated through one of four methods:
Comparison with numerical simulations which encompass larger volumes
such as the Millennium Simulation (e.g., Newman \& Davis~2002;
Somerville et al.~2004; Trenti \& Stiavelli~2008; Moster et al.~2010);
analytically using measurements of the 2 or 3-pt correlation functions
(e.g., Driver et al.~2003); empirically by Monte-Carlo sampling of a
larger survey (e.g., Driver et al.~2005; Hill et al.~2010); or, also
empirically, by Jackknife sampling of the volume in question (e.g.,
Liske et al.~2003). These methods all have strengths and weaknesses
can be laborious to impliment and potentially inconsistent depending
on the method adopted and the assumptions made. For example to
estimate the cosmic variance from numerical simulations requires the
adoption of a numerical simulation (e.g., Springer et al.~2005), 
and either a semi-analytical prescription (Cole et al.~ 2000;
Baugh~2006), or a halo occupation distribution (Berlind \&
Weinberg~2002; Moster et al.~2010) before an appropriate cosmic
variance estimate can be made (e.g., Moster et al.~2010). The
analytical method (e.g., Driver et al.~2003) implicitly assumes
Poisson statistics, radial symmetry, and (because it is parametric)
smoothes over potential 'features' in the underlying distribution
(e.g., BAOs). The empirical method is not always practical if a large
suitable survey does not exist, and Jackknife sampling (where one
divides the sample into many parts and recomputes the value in
question with each part missing in turn) is only capable of revealing
the cosmic variance on scales smaller than the volume in question
(nevertheless a useful indicator as cosmic variance should generally
decrease with increasing scale).

In this paper we aim to use the largest volume survey to date, the
Sloan Digital Sky Survey (SDSS), to empirically determine some
generically useful formulae for estimating the cosmic variance as a
function of survey volume and survey shape. These formulae should also
assist in the design of future surveys where trade offs between area
and depth need to be made. Throughout we adopt a standard cosmology
with the following parameter set: $\Omega_M=0.3,
\Omega_{\Lambda}=0.7, H_o=70$kms$^{-1}$Mpc$^{-1}$ although as our analysis is
based on very local volumes only the value of the adopted Hubble
constant is significant.

\section{The calibration sample}
\begin{figure*}

\vspace{-6.0cm}

\centerline{\psfig{file=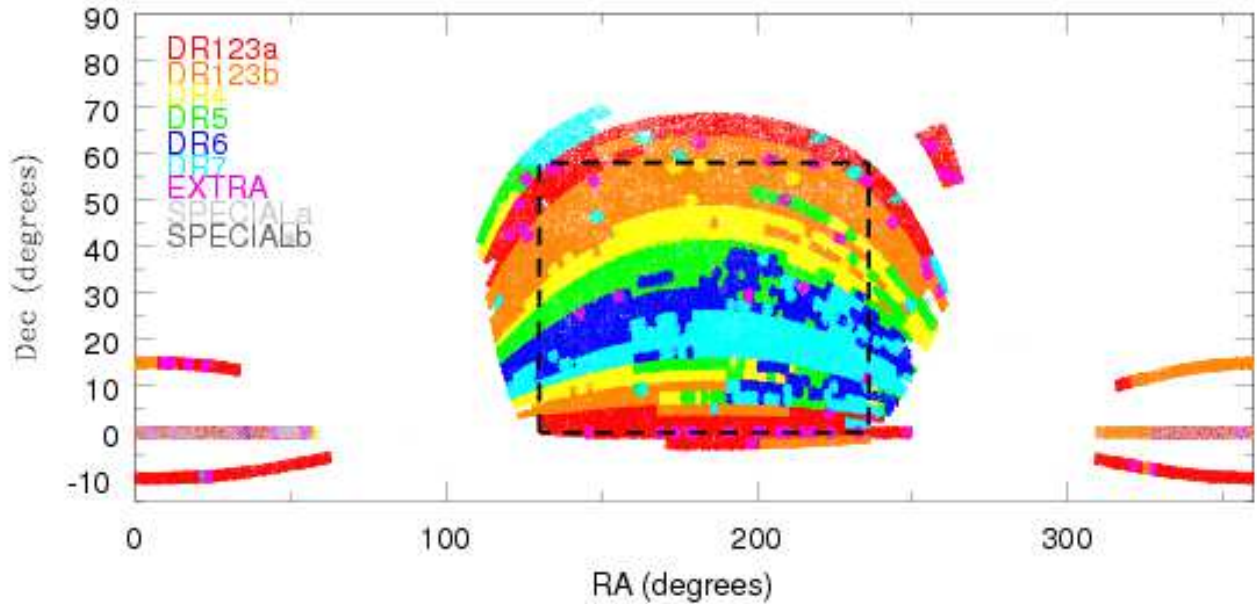,width=\textwidth}}

\centerline{\psfig{file=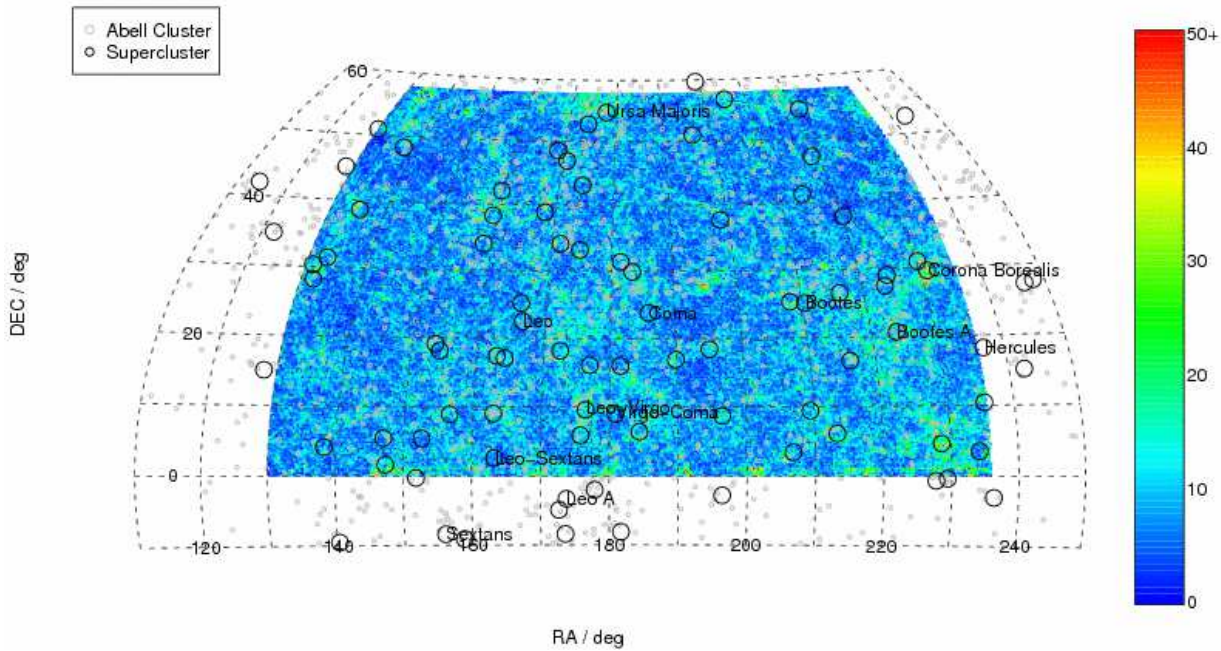,width=\textwidth}}

\caption{({\it upper}) The Sloan Digital Sky Survey shown region by
  region as indicated by the colours and our selected region (black
  dashed line) encompassing 5150 sq.deg. which we adopt for this
  study. ({\it lower}) The distributions of sources on the sky to
  $r<17.77$ mag in 0.1 sq.deg. cells. Large scale structure and cosmic
  variance are clearly present on a variety of scales. Note that the
  spherical sky is shown projected onto a flat geometry resulting in
  cells at higher declination appearing slightly larger. The
    colour bar scale depicts the actual number of galaxies per 0.1
    sq. deg. cell. with $r<17.77$ mag. \label{sdssregion}}
\end{figure*}

The Sloan Digital Sky Survey Data Release 7 (SDSS DR7; Abazajian et
al. 2009) represents the final release of the original main galaxy
survey programme. As our starting point we downloaded the main galaxy
survey spectroscopic compendium
catalogs\footnote{http://www.sdss.org/dr7/products/spectra/getspectra.html
  main galaxy files.}. In total this comprises of the full DR1---7
{\it spectroscopic} catalogues plus files of extra regions (mainly
re-observations of poor quality plates), and two files of special
regions (mainly Stripe 82 observations and leading into SEGUE). From
these data the following fits keyword columns were selected: {\sc ra,
  dec, zfinal, zconffinal, petrocounts, reddening}. From this
extraction only r-band fluxes were selected and these were corrected
for reddening using the {\sc reddening} values provided so that in
what follows, $r_o=r_{\mbox{\sc petrocounts}}-r_{\mbox{\sc
    reddening}}$ in mag units. Fig.~\ref{sdssregion} (upper) shows the
distribution of these data on the sky colour coded according to the
SDSS data release evolution. Estimating the coverage of these data is
non-trivial, a simple method of counting occupied cells of equal area
across the sky suggests it is between 8200 and 8000 sq degrees. The
main difficulty is the complex boundary shape. To simplify the
calculations and minimise boundary concerns we extracted a well
defined sub-region indicated on Fig.~\ref{sdssregion} (upper) as a
black dashed line. This is specified by the coordinates $130.00 <
\alpha(J2000.0) < 236.00, 0.00 < \delta (J2000.0) < 58.00$
representing a pyramidic section of the sky. The area enclosed within
this section is 5150 sq.deg. While holes within this data exist,
mainly small regions due to bright stars, this is to some extent
another kind of cosmic variance (foreground obscuration) operating on
a scale significantly smaller than that in which we are interested ($>
1$ sq.deg). We therefore elect not to use the SDSS mask for four
reasons: firstly speed of computation (in our analysis we will be
extracting many millions of areas and correcting for holes will slow
the code considerably), secondly the area loss is $<3$\%, thirdly it
should not correlate with background structure, and fourthly and perhaps
most importantly the foreground obscuration is a real phenomena.

In our selected sub-region we find 463k galaxies with $r_o<17.77$ mag
of which 415k (90 per cent) have reliable redshifts ({\sc
  zconffinal}$>0.95$).  To assess the uniformity of the data over the
entire region we construct the integrated number-counts to $r_o <
17.77$ mag in 0.1 sq deg cells (as shown in Fig.~\ref{sdssregion}
lower) and perform a bilinear fit to a simple flat projection of this
region (i.e., we interpret Right Ascension and Declination as
Cartesian but keep cell sizes as strictly $0.32 \times 15 \cos(\delta)
0.32$ to generate uniform 0.1 sq deg cells). The fit reveals a weak
gradient in both Right Ascension and Declination, caused by the
presence of a series of superclusters in the lower right corner of
Fig.~\ref{sdssregion} (lower). This illustrates that even the SDSS is
not immune to cosmic variance exhibiting a detectable large scale
structure across the entire target region. Note that the density
gradient extends over a scale of 100 sq deg which at the median
redshift of $\langle z \rangle=0.1$ equates to a ~1Gpc scale
structure. We do not explore the nature of this trend any further but
do note some orthogonal indication of extremely large Gpc-scale
foreground structures arising from WMAP studies (see for example
Hansen, Banday \& G\'orski 2004 and Tangen 2010).

\subsection{Managing spectroscopic incompleteness}
\begin{figure*}
\centerline{\psfig{width=\textwidth,file=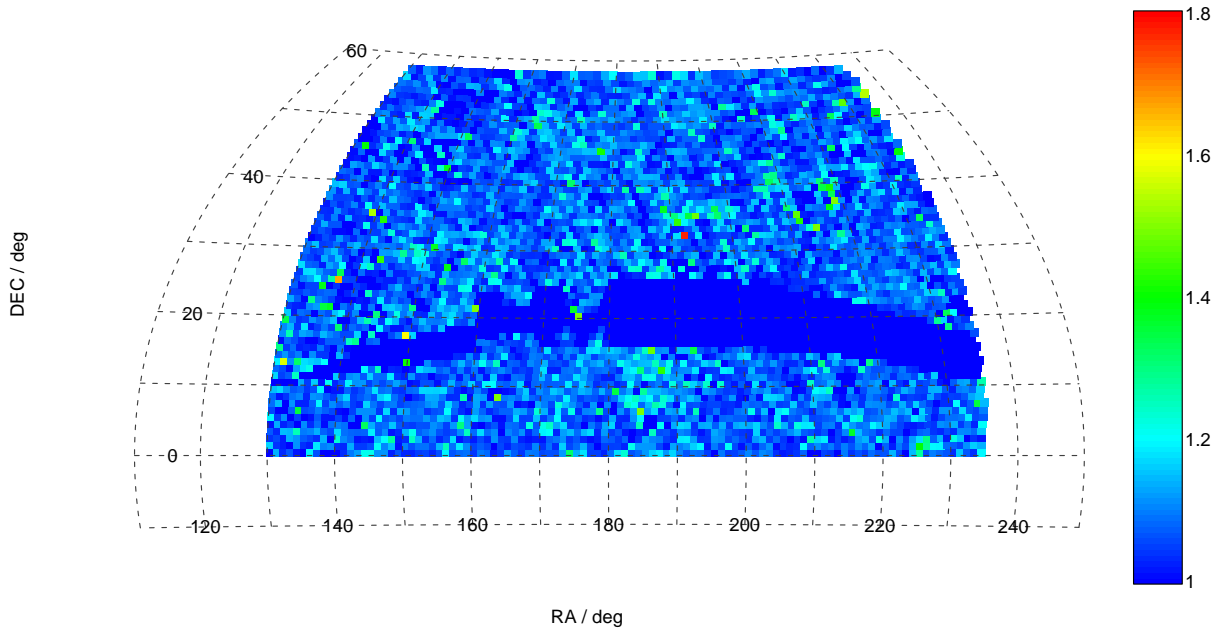}}
\caption{The mean weight (1/completeness, see colour bar scale) within
  1 sq.deg cells across the test region, indicating some minimal
  structure in the spatial completeness but no obvious extended
  regions of poor incompleteness. Note that the few hotspots of
    high incompleteness are aligned with the holes in the SDSS survey
    and represent less than 1 per cent of the data at this
    resolution. \label{incomp}}
\end{figure*}

Spectroscopic incompleteness across the survey is also unlikely to be
uniform but dependent on observing conditions (i.e., extinction,
zenith angle etc) as well as the flux of the object in question. These
cannot be decoupled and must be treated simultaneously. Here we divide
the region into cells of regular sky coverage of 1 sq.deg (containing
on average 75 galaxies). Within each cell we then determine the
completeness as a function of apparent magnitude in 0.25 mag
bins. Each galaxy, $i$, within this cell with a known redshift, $z_i$,
is then assigned a weight, $W_i$. This is the number of galaxies in
some apparent magnitude bin, $N(m_i)$, divided by the number with
known redshift in the same bin, $N(m_i,{\mbox{\sc zconffinal}}>0.95)$
i.e., $W_i=\frac{N(m_i)}{N(m_i,{\mbox{\sc zconffinal}}>0.95)}$.  The
global, or mean, weight for a cell (i.e., integrated over magnitude)
is not particularly meaningful but can be obtained from the average
assigned weight within each cell. Fig.~\ref{incomp} shows the mean
weight in 1 sq.deg cells indicating that the incompleteness is not
strongly clustered but distributed fairly randomly across the region
except for a central swathe of exceptionally high completeness
introduced by virtue of DR7 data.

\subsection{Creating suitable test particles}
\begin{figure}
\centerline{\psfig{width=\columnwidth,file=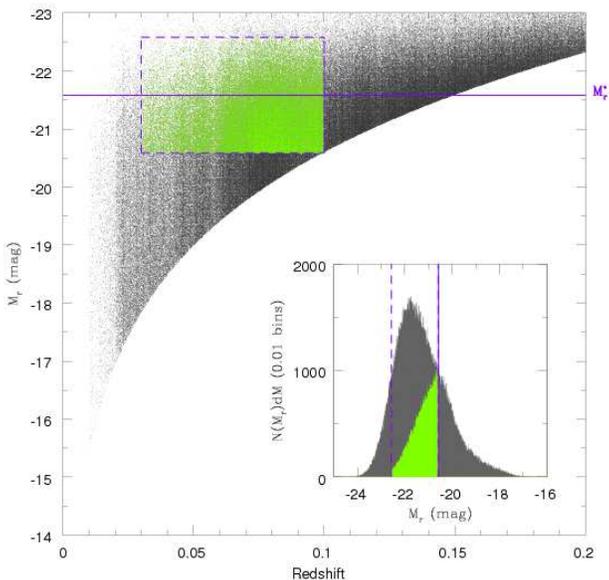}}
\caption{Absolute $r$-band magnitude versus redshift for the data
  within our selected image region. Shown in green is the volume-limited
  region defined by fixed absolute magnitude and redshift limits which
  we adopt. The inset panel shows the data collapsed in redshift. \label{mz}}
\end{figure}

\begin{figure}
\centerline{\psfig{width=\columnwidth,file=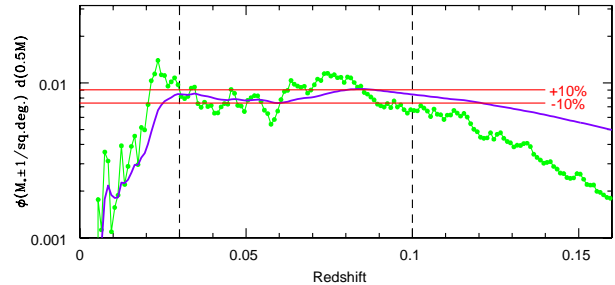}}
\vspace{-4.0cm}
\caption{The density of galaxies with absolute magnitudes $M_r =
  -20.81 \pm 1.00$ mag versus redshift. Both the differential (green
  data) and cumulative (mauve) distributions are shown. The red lines
  show the $\pm 10$ per cent values around the mean value within the
  redshift range indicated by the vertical dashed lines. Large scale
  structure is significant with the known issue of redshift
  incompleteness in the SDSS apparent at low redshift. The turn down
  at higher redshift is where the data become incomplete. The vertical
  lines indicate our selected redshift range for the definition of our
  volume limited sample. \label{cosvar}}
\end{figure}

Having established a sub-region and a mechanism of correcting for the
spectroscopic incompleteness as a function of apparent magnitude and
spatial location we now require a set of test particles to start
exploring the cosmic variance within the test region. We adopt as test
particles the most common galaxy in a locally observed sample, i.e.,
$M^*\pm1.0$ mag galaxies. These are both numerous and readily
detectable to large distances. Selecting brighter systems potentially
introduces additional variance as brighter sources are known to
cluster more strongly (i.e., atypical; c.f., Norberg et al.~2002b),
whereas low luminosity sources would restrict the depth of our volume
and their intrinsic clustering properties are less well known. We
adopt $M_r^*-5\log_{10}(h_{0.7})=-21.58$ mag taken from the recent
$ugrizYJHK$ LF estimates of Hill et al.~(2010). The r-band is used as
this is the filter in which the SDSS main spectroscopic sample is
selected. Fig.~\ref{mz} shows the distribution of absolute magnitude
versus redshift with the systems within our defined volume shown in
green. In deriving the absolute magnitudes we adopted universal $k(z)$
and $e(z)$ corrections ($k(z)=1.263z+0.895z^2-0.566z^3$ and
$e(z)=2.5\log_{10}[(1+z)^{-0.5}]$). The choice of $k(z)$ and $e(z)$
corrections are not critical nor is the assumption of universal rather
than individual corrections as firstly the corrections are small at
low-z ($z<0.1$, see Hill et al.~2010), and secondly we will always
compare cells constructed over identical redshift ranges.  From
Fig.~\ref{mz} we can see that the maximum redshift we're capable of
sampling, due to the SDSS spectroscopic limit, is $z=0.1$.
Fig.~\ref{cosvar} shows the variance along the line of sight by
counting the space-density of test particles (i.e., $M^*\pm1.0$
galaxies) as a function of redshift, both cumulative (mauve curve) and
differential (green data points with errorbars) distributions are
shown. The differential counts show that the SDSS appears to suffer
from an extreme underdensity locally ($z<0.02$) followed by a strong
overdensity at ($z=0.022$) with the cumulative density well behaved
from $z=0.03$ to $z=0.1$ (within 10\% of the mean density in this
range, see red lines on Fig.~\ref{cosvar}) after which incompleteness
(i.e., traditional Malmquist bias) starts to affect the sample
(r.f. Fig.~\ref{mz}). We therefore adopt a z range of 0.03 to 0.10
(minimum/maximum transverse co-moving scale of 2.2/7.3 Mpc/deg) which
contains 100117 test galaxies distributed over an area of 5150 sq.deg
with a co-moving radial length of $291h^{-1}_{0.7}$Mpc, a lookback
interval of $\sim$0.9Gyr, and a volume of $3.7 \times 10^7$
$h^{-3}_{0.7}$Mpc$^{3}$. In the results that follow we will therefore
only be sensitive to cosmic variance on scales below
$10^7h^{-3}_{0.7}$Mpc$^3$, however as cosmic variance is generally
seen to be decreasing the loss of sensitivity to variance on scales
greater than this volume is not expected to be a significant issue. In
what follows the radial co-moving distance is typically comparable or
larger ($\sim 290h^{-1}_{0.7}$Mpc) than the transverse lengths
($<250h^{-1}_{0.7}$Mpc) and is therefore not the dominating dimension
contributing to the cosmic variance. This would not be the case if the
depth was quantised more discreetly, hence for what follows we have
two caveats:

~

\noindent
(1) We are not sensitive to the cosmic variance on volumes larger than
$10^7h^{-3}_{0.7}$Mpc$^{3}$

~

\noindent
(2) We require that the shortest co-moving lengths defining the volume
are tangential to the line-of-sight. In general this will be the case
if the depth interval is $\geq 0.9$Gyr in lookback time.

~

\begin{figure*}

\centerline{\psfig{width=\textwidth,file=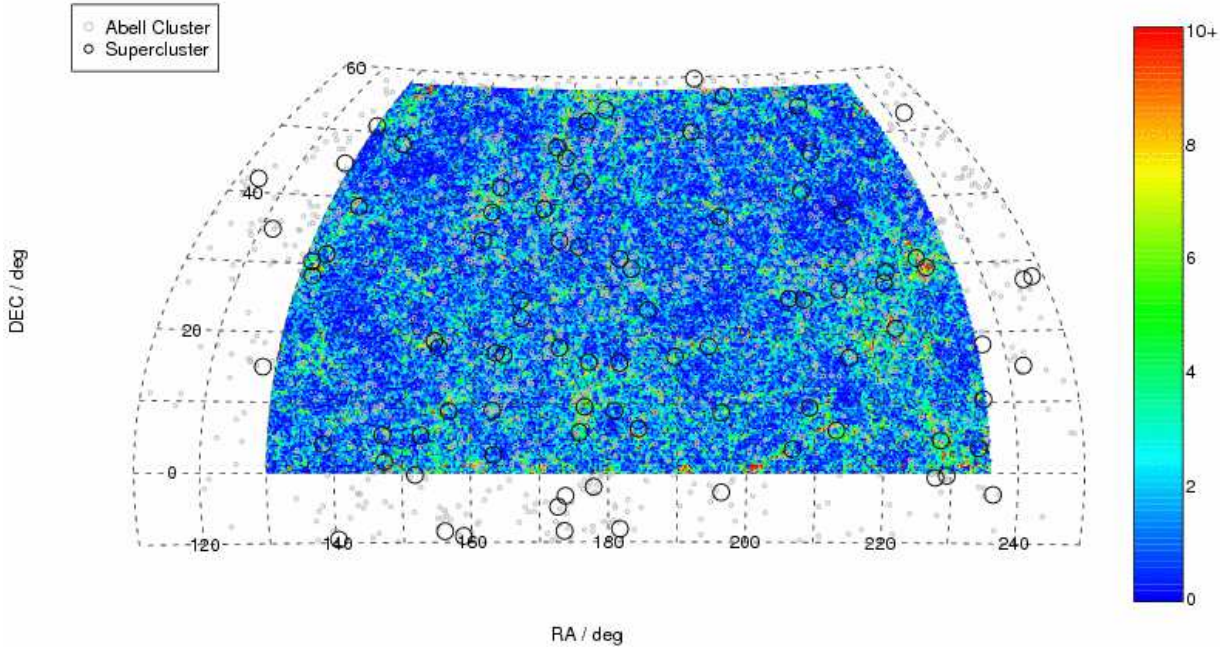}}

\caption{An Aitoff projection of the selected region showing the
  density of $M^*$ systems within the redshift range $0.03 < z <
  0.1$. Clustering and filamentary structure is clearly evident
  across the region. The Abell cluster catalogue and Einisto
  supercluster catalog are overlaid as indicated. \label{testvol}}
\end{figure*}

\begin{figure}
\centerline{\psfig{width=\columnwidth,file=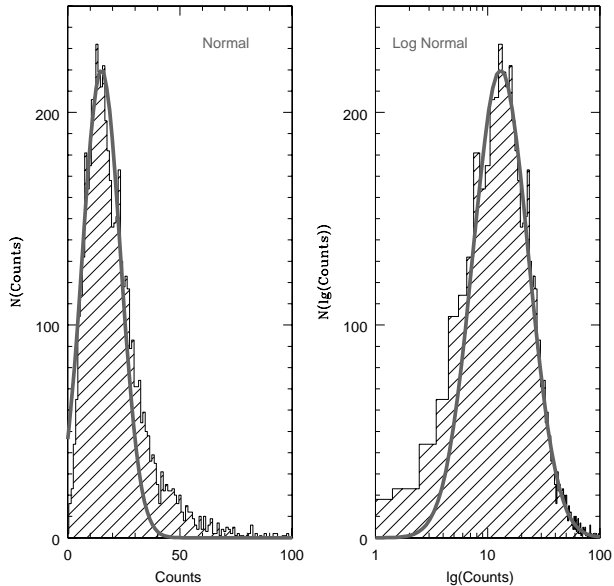}}
\caption{The histogram of counts of test particles in regular 1
  sq.deg. cells across the survey region plotted as both linear (left)
  and logarithmic units (right). Overlaid is a canonical Normal and
  logNormal distribution.  Our data appears to lie somewhere between a
  perfect Normal and a perfect logNormal distribution indicating that
  our choise of statistic, the variance, is an appropriate reference
  measure. \label{gauss}}
\end{figure}

  Finally Fig.~\ref{gauss} shows the histogram of the counts in 1
  sq. deg cells for our test particles indicating a well behaved
  distribution somewhere between an ideal Normal and logNormal
  distribution --- typically in previous studies a Normal distribution
  is assumed and we therefore follow convention. We elect to quantify
  cosmic variance using the simplest measure of standard deviation
  defined as: $\zeta_{\rm Cos. Var.}(\%)=\frac{\sigma_{\rm Var.}}{<N>}
  \times 100$ where $\sigma_{\rm Var.}^2=\frac{\Sigma[<N> -
      N_i]^2}{n}$, $<N>$ is the mean of the counts in cells for that
  particular cell size, and $N_i$ the counts in the $i^{th}$ cell. Note
  that this statistic will by definition include the intrinsic Poisson
  component which in all cases is the minor component (see blue line on
  Fig.~\ref{fig:sample1}).

\section{Results}
Having defined an extensive test region (see Fig.~\ref{testvol})
containing test particles which reflect the underlying local
structures contained in the volume, we can now sample the variance by
repetitively extracting counts in fixed sized cells (truncated
square-based pyramids) at random locations and repeat for increasing
cell sizes. In the next two subsections we firstly explore the basic
variance in square cells from 1 sq.deg to 2048 sq.deg, and secondly
the variance in rectangular cells where the aspect ratio is varied
from 1:1 to 1:128. In the sample extractions which follow we only
sample individual test particles once, resulting in a fully
uncorrelated measure of the cosmic variance but a reduction in the
statistical accuracy for the larger samples and aspect ratios where
fewer independent samples can be constructed.

\subsection{The variance in square regions}
\begin{figure}
\centerline{\psfig{width=\columnwidth,file=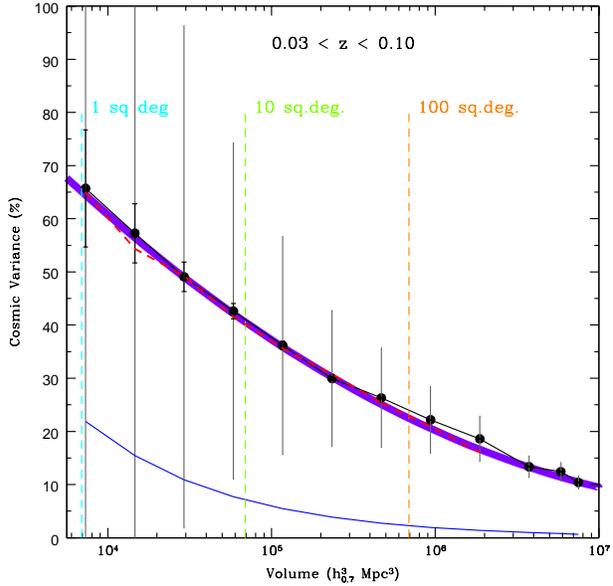}}
\caption{Cosmic variance as a percentage versus volume surveyed. The
  data points show the mean measurements of independent volumes with
  the error on this mean indicated by the errorbars. The grey line
  denotes the range of variances about the mean. The mauve band shows
  our 2nd order polynomial fit to these data. The red line shows the
  empirical result one gets if one uses cuboids rather than pyramids
  for the same volume (see discussion in Section 3.4). The dark blue
    line shows the expected variance from Poisson statistics
    alone. \label{fig:sample1}}
\end{figure}

Fig.~\ref{fig:sample1} shows the (cosmic) variance (data points)
derived from our test region. This is shown as a percentage versus
volume sampled. The range of individual values measured is shown by
the grey bars and the accuracy to which the mean variance is measured
is shown as errorbars. We can see that the variance decreases steadily
from 60 per cent for volumes of $10^4h^{-3}_{0.7}$Mpc$^3$ to 10 per
cent at our sampling limit of $10^7h^{-3}_{0.7}$Mpc$^{3}$. We fit a
simple second order polynomial to the data and find a good ($< \pm 1$
percent) fit given by:
\begin{equation}
\zeta_{\rm Cos. Var.}(\%)=219.7-52.4\log_{10}[V]+3.21(\log_{10}[V])^2
\end{equation}
where $\zeta_{\rm Cos. Var}$ represents the cosmic variance for a
volume, $V$, as a percentage (i.e., $\frac{100\Delta N}{N}$). This
expression can be used to provide a robust estimate of the cosmic variance for any
given square shaped $z<0.1$ survey given the total volume sampled.

\subsection{Contiguous versus sparse sampling}
\begin{figure}
\centerline{\psfig{width=\columnwidth,file=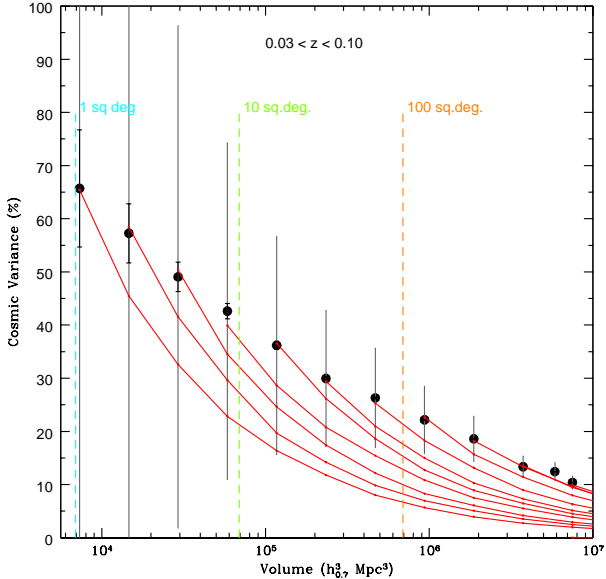}}
\caption{As for Fig.~\ref{fig:sample1} but now showing the efficiency
  of sparse sampling (red lines). The red lines scale exactly as one
  would expect from dividing by root-n where n is the number of
  independent observations of that specific resolution. Note that the
  origin of the red curves may be offset from the large data points
  due to uncertainty introduced by the random sampling of the parent
  distribution. \label{fig:sample2}}
\end{figure}

Given the scale of cosmic variance it is worth asking whether it can
be more easily overcome by multiple independent sight-lines rather
than a single contiguous survey. To explore this we build up a larger
survey by combining $N$ independent regions each of $x$ sq.deg to
obtain a survey of area $Nx$. Fig.~\ref{fig:sample2} shows the results
as red lines originating from the building block area ($x$).
Unsurprisingly the cosmic variance of a larger survey is significantly
reduced if comprised of multiple smaller blocks rather than a single
contiguous survey. For example for a survey of 32 sq.deg constructed
from multiple 1 sq deg. blocks the cosmic variance is reduced from 31
per cent to 11 per cent. The cosmic variance empirically decreases
with multiple sight-lines by $\sqrt{N}$, as one would expect if the
sightlines are indeed decoupled. This holds regardless of the base
survey area (i.e., independent of $x$). Hence for multiple sight-lines
Eqn.~1 can now be modified to:
\begin{equation}
\zeta_{\rm Cos. Var.}(\%)=(219.7-52.4\log_{10}[V]+3.21(\log_{10}[V])^2)/\sqrt{N}
\end{equation}
where $N$ is the number of independent sight lines each of volume $V$.
Similarly the cosmic variance can be determined for independent
regions of differing area and combined assuming Poisson statistics as
long as the blocks are fully independent and non-contiguous.

\begin{figure}
\centerline{\psfig{width=\columnwidth,file=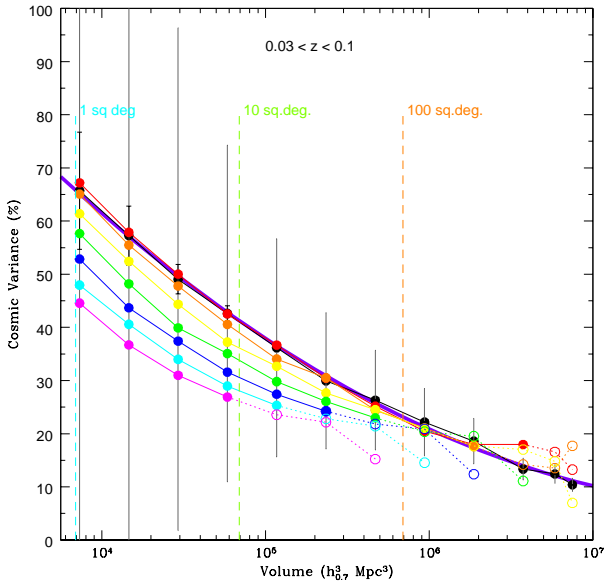}}
\caption{As for Fig.~\ref{fig:sample1} but now including the impact of
  on-sky area aspect ratio from 2:1 (red), 4:1 (orange), 8:1 (yellow),
  16:1 (green), 32:1 (blue), 64:1 (cyan), 128:1 (magenta). Variance is
  significantly reduced when extreme rectangular areas are
  sampled. Note open symbols and dashed lines are used when the
  largest tangential lengths exceeds the radial length of the survey
  region \label{fig:sample3}}
\end{figure}

\subsection{Dependence on aspect ratio}
Survey shape and in particular the aspect ratio, or window function,
is likely to impact on the cosmic variance. Not all surveys will be
square/circular and may have significantly different dimensions in
length and width. For example the Millennium Galaxy Catalogue (Liske
et al., 2003; Driver et al 2005) has a width of 0.5 deg and a length
of 75 sq deg giving an aspect ratio of 1:150. Similarly the GAMA
survey (Driver et al.~2009; Baldry et al.~2010) consists of 3 chunks
of 4 deg by 12 deg regions for an aspect ratio of 1:3. This aspect
ratio, when taken to the extreme, can help significantly to reduce
cosmic variance. Observationally long thin strips are often easier to
observe because of the Earth's rotation and are more robust to cosmic
variance however for the same reasons are less suitable for studies of
large scale structure. Fig.~\ref{fig:sample3} shows the outcome of
modifying the aspect ratio by reproducing Fig.~\ref{fig:sample1} for
various aspect ratios ranging from 1:1 to 1:128. The cosmic variance
follows an almost identical trend as for Fig.~\ref{fig:sample1} but
offset in amplitude. Note that the data becomes noisier for higher
aspect ratios because of the limited number of independent samplings
possible. For very large volumes the aspect range no longer appears to
provide a gain in cosmic variance, this is because the longer
tangential length approaches and exceeds the depth of the test volume
resulting in a cap to the gain in cosmic variance as any volumes
cosmic variance is always dominated by the two shortest lengths. In
these cases the data points are shown as open symbols and the
connecting lines as dotted. Finally we can incorporate the aspect
ratio into Eqn.~2 by simply allowing the amplitude to vary, thus:
\begin{eqnarray}
\zeta_{\rm Cos. Var.}(\%) & = & (1.00-0.03(\sqrt{X-1})) \nonumber \\
& & \times (219.7-52.4\log_{10}[V] \nonumber \\
& & +3.21(\log_{10}[V])^2)/\sqrt{N}
\end{eqnarray}
where $X$ is the aspect ratio, e.g., 128 for 1:128. Eqn.~3 now
provides a robust estimate of the cosmic variance for the interval
$z<0.1$ for almost any survey in terms of the sampling volume, $V$,
the aspect ratio of the survey window, $X$, and the number of
independent volumes, $N$.

\subsection{Generalising over all redshift for any survey}
Extrapolating the current method beyond $z \approx 0.1$ becomes
non-trivial for a number of reasons: firstly, and foremost, the
clustering signature of the population is evolving, with the galaxy
population expected to be less clustered towards higher redshift,
secondly one needs to consider the three dimensional volume shape,
(i.e., in the earlier section we kept the redshift baseline constant
at $\Delta z \sim 0.07$ which equates to a physical co-moving distance
of $\sim 291h^{-1}_{0.7}$ Mpc). The first of these issues cannot be
addressed using the SDSS and, as no suitable dataset exists at
higher-z, it is currently empirically intractable. However one can
adopt the values from Eqn.~3 as a robust upper limit. The issue of
survey shape is also itself problematic in two ways: firstly as one
moves to higher-z, for a fixed survey window, the volume becomes less
conic/pyramidic and more cylindrical/cuboid due to the tendency
towards a nearly constant angular-diameter-transverse length relation;
secondly the freedom allowed by modifying all three dimensions of the
sampling volume makes the derivation of a direct empirical expression
valid beyond our maximum length impossible. In a future paper
(Robotham \& Driver, in prep.) we will address the first of these by
providing an online Cosmic Variance calculator which allows the user
to specify the precise dimensions of their survey cuboid upto
approximately $300\times 300\times 250h^{-3}_{0.7}$Mpc$^3$
volumes. However this may still not cover the very deep pencil beam
surveys where one might wish to bin results over radial co-moving
lengths greater than $300h^{-1}_{0.7}$Mpc. However we can make two
relatively simple assumptions to provide an analytical workaround.

~

\noindent
(1) Over scales greater than $\sim250h^{-1}_{0.7}$Mpc one expects no correlation
of structure (although note our apparent detection of weak structure
over a 1Gpc linear scale across the entire SDSS). Certainly if one
axis is significantly greater than the other two one expects the
cosmic variance across the longer dimension to contribute least
to the total variance. Hence the cosmic variance in long cuboids
should scale according to Poisson statistics if the long radial length is
increased/decreased over a range $>200h^{-1}_{0.7}$ Mpc. i.e., a volume
of $5.3 \times 5.3 \times 500.0h^{-3}_{0.7}$Mpc$^3$ should have $\sqrt{2}$ less variance
than that defined by $5.3 \times 5.3 \times 250.0h^{-3}_{0.7}$Mpc$^{3}$.

~

\noindent
(2) We equate any survey volume to a cuboid where we preserve the total
volume, $\Delta z$ range, and aspect ratio and derive the appropriate
transverse lengths. Thus our $7320h^{-1}_{0.7}$Mpc volume sampled by 1
sq.deg to $0.03 < z < 0.1$ can be equated to a cuboid of dimensions
$5.015 \times 5.015 \times 291h^{-3}_{0.7}$Mpc$^3$.

~

The first of these assumptions is reasonable assuming the radial
length is always the greater of the three cuboid sides. The second can
be tested by extracting equal area and equal radial length square
based pyramids and cones. Adopting a constant radial length of
291$h^{-1}_{0.7}$Mpc we reproduce the initial results shown on
Fig.~\ref{fig:sample1} by now sampling our volume with cuboids (red
line) rather than the original square-based pyramids (data points). To
derive our cuboids we fix the total volume, the $\Delta z$ range and
derive the required transverse co-moving lengths. The red curve
closely follows the original data points indicating no correction is
required for the change in geometric shape.

Given these two caveats one can now trivially determine an {\it
  approximate} cosmic variance for any survey volume.  We
achieve this by replacing the survey volume, $V$, with the product of
the median redshift transverse lengths (for the survey bin in
question), $A$ \& $B$, and a radial depth $C$ all expressed in
$h^{-1}_{0.7}$Mpc, then, assuming that $C> 250h^{-1}_{0.7}$Mpc, we find from
Eqn.~3 and the caveats above that:
\begin{eqnarray}
\zeta_{\rm Cos. Var.}(\%) & = & (1.00-0.03(\sqrt{(A/B)-1})) \nonumber \\
& & \times (219.7-52.4\log_{10}[A.B.291.0] \nonumber \\
& & +3.21(\log_{10}[A.B.291.0])^2) \nonumber \\
& & /(\sqrt{N.\frac{C}{291.0}})
\end{eqnarray}
Note that the ratio of the transverse lengths ($A/B$) at the median
redshift has replaced the Aspect ratio ($X$) in Eqn.~3. For
conic/cylindrical surveys one can replace $A$ and $B$ with
$\sqrt(\pi)R$ where $R$ is the transverse radius of the survey at the
median redshift. This equation is strictly valid over the range
$3 \times 10^{3}$ to $1 \times 10^{7} h_{0.7}^{-3}$Mpc$^3$


\subsection{Comparison to predictions from simulations}
In a recent study Moster et al.~(2010) derived cosmic variance values
from a purely numerical/analytical route using simulations coupled
with the adoption of a Halo Mass Function. We reproduce on Table.~1
columns from their Table.~5 for galaxies of mass
$10^{10.75}$M$_{\odot}$, the approximate turn-over point of the
stellar mass function (see Baldry, Glazebrook \& Driver~2008), and
compare to our predictions using Eqn.~4 for the GOODs, GEMs, and
COSMOS surveys. In general the values shown in Table.~3 paint a
remarkably consistent picture of the cosmic variance with estimates
based on the two entirely distinct methods producing remarkably
consistent values for $z<2.39$. One interesting offset is the tendency
for the empirical method to produce lower cosmic variance values at
higher redshift. While this initially appears counter-intuitive it can
be explained in the context that while general clustering {\it
  decreases} as a function of redshift it presumably {\it increases}
for fixed stellar mass. In other words the $10^{10.75}M_{\odot}$
systems at high-z are destined to become the highly clustered
superluminous population at low redshift. This is borne out if we
compare our cosmic variance values to the lower mass values of Moster
et al. at higher redshift. For example for the cosmic variance in a
GOODS field in redshift interval 3.58 to 4.00 for stellar masses of
$10^{9.25}M_{\odot}$ Moster et al find $\sigma_{\rm RMS}=0.296$
consistent with our value of 30.6\%.  We therefore conclude that our
method continues to provide a reasonable estimate of the $M^*$ cosmic
variance at any redshift and that the discrepancy between Moster et
al. and our work at high redshift is fully explained by the change in
$M^*$ stellar mass as a function of redshift. One should therefore use
our formulae if the $M^*$ point is sampled or the Moster et
al. formulae for know stellar mass ranges.

\begin{table*}
\caption{Comparison of cosmic variance estimations from this work (Col.
  4) with Moster et al.~(2010; Col. 3) for surveys and redshift ranges
  indicated in Cols.~1\&2.}
\begin{tabular}{c|c|c|c} \hline \hline
Survey & z    & Moster et al.~(2010) & This work \\
       & range & $\sigma_{\rm RMS}$ for $10^{10.75}M_{\odot}$ & (\%) \\ \hline
GOODS ($10' \times 16')$ & 0.00---1.12 & 0.126 & 15.5\% \\
      & 1.12---1.58 & 0.194 & 23.0\% \\
      & 1.58---1.99 & 0.241 & 25.1\% \\
      & 1.99---2.39 & 0.295 & 26.5\% \\
      & 2.39---2.78 & 0.365 & 28.0\% \\
      & 2.78---3.17 & 0.446 & 29.2\% \\
      & 3.17---3.58 & 0.534 & 29.7\% \\
      & 3.58---4.00 & 0.647 & 30.6\% \\ \hline
GEMS ($28' \times 28')$ & 0.00---1.12 & 0.098 & 11.3 \\
      & 1.12---1.58 & 0.140 & 16.0\% \\
      & 1.58---1.99 & 0.169 & 17.3\% \\
      & 1.99---2.39 & 0.203 & 18.1\% \\
      & 2.39---2.78 & 0.247 & 19.0\% \\
      & 2.78---3.17 & 0.299 & 19.7\% \\
      & 3.17---3.58 & 0.354 & 20.0\% \\
      & 3.58---4.00 & 0.425 & 20.5\% \\ \hline
COSMOS ($84' \times 84')$ & 0.00---1.12 & 0.057 & 6.5\% \\
      & 1.12---1.58 & 0.069 & 8.5\% \\
      & 1.58---1.99 & 0.080 & 9.0\% \\
      & 1.99---2.39 & 0.093 & 9.3\% \\
      & 2.39---2.78 & 0.110 & 9.7\% \\
      & 2.78---3.17 & 0.131 & 10.0\% \\
      & 3.17---3.58 & 0.153 & 10.1\% \\
      & 3.58---4.00 & 0.181 & 10.4\% \\ \hline
\end{tabular}
\end{table*}

\section{Cosmic variance values for specific surveys.}
The main purpose of this paper is to derive a simple credible path to
empirically based cosmic variance estimates for recent, ongoing, and
upcoming surveys. Table.~2 shows a variety of cosmic variance
estimates based on Eqn.~4. The main conclusion is the need for
multiple independent sightlines to reduce the impact of cosmic
variance. In particular any deep ASKAP survey should have a minimum of
2 ultra-deep fields and any deep MeerKAT survey should have a minimum
of 10 ultra-deep fields to probe the HI universe in $0.1 \Delta z$
intervals and keep cosmic variance below 10\%. Likewise approximately
100 HST ACS or 10 HST WFC3 ultra-deep fields are required to keep
cosmic variance below 10\% in $\Delta z$ intervals of 1. Finally we
note that the GAMA, VVDS and zCOSMOS surveys all have cosmic variance
below the 10\% level in intervals of $\Delta z$ of 0.1, 0.25 and 0.5
respectively.

\begin{table*}
\caption{Cosmic variance values for various ongoing and planned
  surveys defined from Eqn.~4.}
\begin{tabular}{c|c|c|c|c|c|c} \hline \hline
Survey & Redshift & Area & Aspect & Cos. Var. per & Number of & Final \\ 
Name & range & sq.deg. & ratio & pointing & pointings & Cos. Var \\ \hline
MGC & 0.0 --- 0.1 & 30 & 1:128 & 19\% & 1 & 19\% \\
MGC & 0.0 --- 0.2 & 30 & 1:128 & 10\% & 1 & 10\% \\ \hline
GAMA I & 0.0 --- 0.1 & 48 & 1:3 & 24\% & 3 & 14\% \\
GAMA I & 0.1 --- 0.2 & 48 & 1:3 & 15\% & 3 & 8\% \\
GAMA I & 0.2 --- 0.3 & 48 & 1:3 & 11\% & 3 & 6\% \\
GAMA I & 0.3 --- 0.4 & 48 & 1:3 & 9\% & 3 & 5\% \\ 
GAMA I & 0.4 --- 0.5 & 48 & 1:3 & 8\% & 3 & 5\% \\
 GAMA I & 0.0 --- 0.5 & 48 & 1:3 & 5\% & 3 & 3\% \\ \hline
ASKAP  & 0.0 --- 0.1 & 36 & 1:1 & 27\% & 2 & 19\% \\
ASKAP  & 0.1 --- 0.2 & 36 & 1:1 & 17\% & 2 & 12\% \\
ASKAP  & 0.2 --- 0.3 & 36 & 1:1 & 13\% & 2 & 9\% \\
ASKAP  & 0.3 --- 0.4 & 36 & 1:1 & 11\% & 2 & 8\% \\
ASKAP  & 0.4 --- 0.5 & 36 & 1:1 & 10\% & 2 & 7\% \\ \hline
MeerKAT & 0.0 --- 0.1 & 1 & 1:1 & 58\% & 10 & 18\% \\ 
MeerKAT  & 0.1 --- 0.2 & 1 & 1:1 & 41\% & 10 & 13\% \\
MeerKAT  & 0.2 --- 0.3 & 1 & 1:1 & 34\% & 10 & 11\% \\
MeerKAT  & 0.3 --- 0.4 & 1 & 1:1 & 30\% & 10 & 10\% \\
MeerKAT  & 0.4 --- 0.5 & 1 & 1:1 & 28\% & 10 & 9\% \\
MeerKAT  & 0.5 --- 0.6 & 1 & 1:1 & 26\% & 10 & 8\% \\
MeerKAT  & 0.6 --- 0.7 & 1 & 1:1 & 25\% & 10 & 8\% \\
MeerKAT  & 0.7 --- 0.8 & 1 & 1:1 & 24\% & 10 & 8\% \\
MeerKAT  & 0.8 --- 0.9 & 1 & 1:1 & 24\% & 10 & 7\% \\
MeerKAT  & 0.9 --- 1.0 & 1 & 1:1 & 23\% & 10 & 7\% \\ \hline
VVDS & 0.00 --- 0.50 & 4 & 1:1 & 10\% & 4 & 5\% \\
VVDS & 0.50 --- 0.75 & 4 & 1:1 & 10\% & 4 & 5\% \\
VVDS & 0.75 --- 1.00 & 4 & 1:1 & 10\% & 4 & 5\% \\
VVDS & 1.00 --- 1.25 & 4 & 1:1 & 9\% & 4 & 5\% \\
VVDS & 1.25 --- 1.50 & 4 & 1:1 & 9\% & 4 & 5\% \\
VVDS & 1.50 --- 1.75 & 4 & 1:1 & 9\% & 4 & 5\% \\ 
VVDS & 1.75 --- 2.00 & 4 & 1:1 & 9\% & 4 & 5\% \\ \hline
zCOSMOS & 0.0 --- 0.5 & 2 & 1:1 & 12\% & 1 & 12\% \\
zCOSMOS & 0.5 --- 1.0 & 2 & 1:1 & 9\% & 1 & 9\% \\
zCOSMOS & 1.0 --- 1.5 & 2 & 1:1 & 8\% & 1 & 8\% \\
zCOSMOS & 1.5 --- 2.0 & 2 & 1:1 & 8\% & 1 & 8\% \\
zCOSMOS & 2.0 --- 2.5 & 2 & 1:1 & 8\% & 1 & 8\% \\ \hline 
HST ACS  & 1.0 --- 1.5 & 4.8E-05 & 1:1 & 71\% & 10 & 22\% \\ 
HST ACS  & 1.5 --- 2.0 & 4.8E-05 & 1:1 & 77\% & 10 & 24\% \\
HST ACS  & 2.0 --- 3.0 & 4.8E-05 & 1:1 & 61\% & 10 & 19\% \\
HST ACS  & 3.0 --- 4.0 & 4.8E-05 & 1:1 & 70\% & 10 & 22\% \\
HST ACS  & 4.0 --- 5.0 & 4.8E-05 & 1:1 & 79\% & 10 & 25\% \\
HST ACS  & 5.0 --- 6.0 & 4.8E-05 & 1:1 & 87\% & 10 & 28\% \\
HST ACS  & 6.0 --- 7.0 & 4.8E-05 & 1:1 & 96\% & 10 & 30\% \\
HST ACS  & 7.0 --- 20.0 & 4.8E-05 & 1:1 & 39\% & 10 & 12\% \\ \hline
HST WFC3  & 1.0 --- 1.5 & 1.3E-03 & 1:1.1 & 44\% & 10 & 14\% \\
HST WFC3  & 1.5 --- 2.0 & 1.3E-03 & 1:1.1 & 46\% & 10 & 15\% \\
HST WFC3  & 2.0 --- 3.0 & 1.3E-03 & 1:1.1 & 36\% & 10 & 11\% \\
HST WFC3  & 3.0 --- 4.0 & 1.3E-03 & 1:1.1 & 41\% & 10 & 13\% \\ 
HST WFC3  & 4.0 --- 5.0 & 1.3E-03 & 1:1.1 & 46\% & 10 & 15\% \\ 
HST WFC3  & 5.0 --- 6.0 & 1.3E-03 & 1:1.1 & 51\% & 10 & 16\%\\ 
HST WFC3  & 6.0 --- 7.0 & 1.3E-03 & 1:1.1 & 55\% & 10 & 17\%\\ 
HST WFC3  & 7.0 --- 20.0 & 1.3E-03 & 1:1.1 & 22\% & 10 & 7\% \\ \hline
\end{tabular}
\end{table*}

\section{Conclusions}
We have derived a simple empirical expression for calculating cosmic
variance for almost any extragalactic survey. The results are entirely
empirical and based on resampling the SDSS DR7. The resulting
equations agree extremely well with the recent numerical results by
Moster et al.~(2010). The two resulting equations provide corrections
for $z<0.1$ robustly and for $z>0.1$ under the following caveats:

~

\noindent
(1) Te derived cosmic variance is for $M^* \pm 1$ mag population only and
assumed not to evolve with lookback time  -- this is clearly
incompatible with our understanding of the evolution of structure and
hence beyond $z \sim 1$ the derived values should be taken as
indicative only.

~

\noindent
(2) That above $250h^{-1}_{0.7}$Mpc cosmic variance scales with radial
co-moving length according to Poisson statistics.

~

The two equations are then used to determine cosmic variance values
for a number of recent, ongoing and planned surveys.

\section*{Acknowledgments}
SPD thanks the University of Western Australia and the International
Centre for Radio Astronomy Research (ICRAR) for financial support and
to Profs Peter Quinn and Lister Staveley Smith, and Dr Martin Meyer
for stimulating discussions on this topic during his Sabbatical stay.

Funding for the SDSS and SDSS-II has been provided by the Alfred
P. Sloan Foundation, the Participating Institutions, the National
Science Foundation, the U.S. Department of Energy, the National
Aeronautics and Space Administration, the Japanese Monbukagakusho, the
Max Planck Society, and the Higher Education Funding Council for
England. The SDSS Web Site is http://www.sdss.org/.

The SDSS is managed by the Astrophysical Research Consortium for the
Participating Institutions. The Participating Institutions are the
American Museum of Natural History, Astrophysical Institute Potsdam,
University of Basel, University of Cambridge, Case Western Reserve
University, University of Chicago, Drexel University, Fermilab, the
Institute for Advanced Study, the Japan Participation Group, Johns
Hopkins University, the Joint Institute for Nuclear Astrophysics, the
Kavli Institute for Particle Astrophysics and Cosmology, the Korean
Scientist Group, the Chinese Academy of Sciences (LAMOST), Los Alamos
National Laboratory, the Max-Planck-Institute for Astronomy (MPIA),
the Max-Planck-Institute for Astrophysics (MPA), New Mexico State
University, Ohio State University, University of Pittsburgh,
University of Portsmouth, Princeton University, the United States
Naval Observatory, and the University of Washington.

\section*{References}

\noindent Abazajian K.N., et al., 2009, ApJS, 182, 543

\noindent Baldry I. K., et al., 2010, MNRAS, 404, 86

\noindent Baldry I.K., Glazebrook K., Driver S.P., 2008, MNRAS, 388, 945

\noindent Baugh C., 2006, RPPh, 69, 3101

\noindent Berlind A.A., Weinberg D.H., 2002, ApJ, 575, 587

\noindent Cole S., Lacey C. G. Baugh C. M., Frenk C.S., 2000, MNRAS, 319, 168  

\noindent Davies M, Efstathiou G., Frenk C.S., White S.D.M., 1985, ApJ, 292, 371

\noindent Driver S.P., Odewahn S.C., Echevarria L., Cohen S.H., Windhorst R.A., Phillipps S., 2003, AJ, 126, 2662

\noindent Driver S.P., Liske J., Cross N.J.G., De Propris R. Allen P.D., 2005, MNRAS, 360, 81

\noindent Driver S.P., et al., 2009, A\&G, 50, 12

\noindent Hill D., Driver S.P., Cameron E., Cross N.J.G., Liske J., Robotham A., 2010, MNRAS, 404, 1215

\noindent Hansen F.K., Banday A.J., G\'orski K.M., 2004, 354, 641

\noindent Hopkins A.M., Beacom J.F., 2006, ApJ, 651, 142

\noindent Liske J., Lemon D.J., Driver S.P., Cross N.J.G., Couch W.J., 2003, MNRAS, 344, 307

\noindent Moster B.P., Somerville R.S., Newman J.A., Rix, H-W, 2010, ApJ, submitted (astro-ph/1001.1737)

\noindent Newman J.A., Davis M., 2002, ApJ, 564, 567

\noindent Norberg P., et al., 2002a, MNRAS, 336, 907

\noindent Norberg P., et al., 2002b, MNRAS, 332, 827

\noindent Somerville R.S., Lee K., Ferguson H.C., Gardner J.P., Moustakas LA., Giavalisco M., 2004, ApJ, 600, 171 

\noindent Springel V et al., 2005, Nature, 435, 629

\noindent Szapudi I., Columbi S., 1996, ApJ, 470, 131

\noindent Tangen K., 2010, submitted, (astro-ph/0910.4164)

\noindent Trenti M., Stiavelli M., 2008, ApJ, 676, 767

\noindent Wilkins S.M. Trentham N., Hopkins A.M., MNRAS, 385, 687 

\noindent Zwaan M., Meyer M., Staveley-Smith L., Webster R., 2005, MNRAS, 359, 30

\label{lastpage}

\end{document}